# Semiautomatic Simplification

Gong Li
gongli@cs.ualberta.ca
Dept. Computing Science
Univ. Alberta
Edmonton, Alberta
CANADA T6G 2H1

Benjamin Watson
watson@nwu.edu
Dept. Computer Science
Northwestern Univ.
1890 Maple Ave
Evanston, IL 60201 USA

**ABSTRACT**

We present `semisimp`, a tool for semiautomatic simplification of three dimensional polygonal models. Existing automatic simplification technology is quite mature, but is not sensitive to the heightened importance of distinct semantic model regions such as faces and limbs, nor to simplification constraints imposed by model usage such as animation. `semisimp` allows users to preserve such regions by intervening in the simplification process. Users can manipulate the order in which basic simplifications are applied to redistribute model detail, improve the simplified models themselves by repositioning vertices with propagation to neighboring levels of detail, and adjust the hierarchical partitioning of the model surface to segment simplification and improve control of reordering and position propagation.

**ACM Category and Subject Descriptor:** I.3.5 [Computer Graphics] Computational Geometry and Object Modeling - *hierarchy and geometric transformations*

## 1 MOTIVATION

Researchers have for several years recognized the need to reduce the complexity of polygonal models, while at the same time preserving their appearance and meaning. This has led to a large and useful body of research on automating this process [6]. However, particularly when models are simplified to a few thousands of vertices or less, automatic algorithms show their limitations. Among these are:

- *Semantic blurring.* Automatic algorithms use fairly simple error measures to guide simplification. These measures cannot adequately gauge the impact of changes to the model that may blur or completely remove features containing high level perceptual or semantic meaning.

- *Functional blurring.* Models may be put to uses that cannot be divined from their geometry, topology or attributes. Such uses may impose certain constraints on the simplification process. For example, when models are animated with articulated skeletons, simplification across the joints of the skeleton can be extremely problematic.

- *Inadequate control.* It may be possible for users to embed additional information into the model to enable automatic simplification to avoid semantic and functional blurring. Nevertheless, users would ultimately like to have this control themselves, particularly when simplified models have a manageable number of faces. Current algorithms offer only the most indirect sort of control through command line parameters.

These limitations have led us to create the semiautomatic simplification tool `semisimp`, which provides control of the simplification process and enables avoidance of both semantic and functional blurring.

## 2 OVERVIEW

`semisimp` is a unique synthesis of simplification and multiresolution modeling functions, emphasizing the improvement of aggressively simplified models. It begins by accepting a fully detailed model as input and applying an automatic simplification algorithm to construct a simplification hierarchy. Users can then edit and improve this hierarchy for their target application in three ways:

- *Order manipulation.* Users can adjust the distribution of detail on simplified models by changing the order in which model regions are simplified. This is accomplished through matching changes in the order in which the simplification hierarchy is traversed.

- *Geometric manipulation.* Users can improve the positions of vertices in simplified models. These improvements can be automatically propagated to both simpler and more detailed models. Propagation to more detailed models can be attenuated to preserve the shape of the original model.

- *Hierarchy manipulation.* Users can halt the simplification, modify the partitioning of the original model described by the partial simplification hierarchy to match semantics and intended model use, and continue simplification in a segmented fashion. Since both geometric and order manipulation operate in the context of the simplification hierarchy, their effectiveness is greatly increased.

Having improved the simplification hierarchy, users can extract discrete levels of detail from this hierarchy, or output the hierarchy itself for use in applications that dynamically adjust level of detail [10,14,19].

## 3 CONTRIBUTIONS

These combined functions offer users a new degree of control over model simplification, and enable users to manually improve heavily simplified models with automatic assistance. By repartitioning the model, redistributing detail, and improving simplified geometry, users can effectively reduce semantic and functional blurring. Novel components of the system include:

- Propagation of repositioned vertices to neighboring levels of detail in a simplification hierarchy, with attenuation of this propagation to preserve the shape of the original model.

- Manual restructuring of the simplification hierarchy, allowing segmented simplification of the model and improved control of simplification order and geometry propagation.

In the remainder of this paper, we review related work (section 4), the functions of semisimp (section 5), and give examples of semisimp's usage (section 6).

## 4 RELATED RESEARCH

Automatic model simplification is by now a mature area of research, with dozens of very effective algorithms. We will not attempt a comprehensive review of these algorithms, but will instead focus on those algorithms of particular relevance in our research context. For an excellent comprehensive review, see [6].

semisimp can be used with a large number of automatic simplification algorithms that implement what we call the *greedy search* paradigm. These algorithms work by identifying a number of possible *primitive simplifications* and choosing among them, using an iterative, greedy search algorithm. This is achieved by estimating the error each primitive simplification would introduce using a *simplification error measure*, and inserting it into a *primitive simplification queue* that is sorted by error. During each iteration, the primitive simplification that would introduce the least error is removed from the front of the queue, the complex surface it removes is approximated using a *simplification filter*, and the queue is updated to include the new primitive simplifications possible in the affected model neighborhood. The history of this process when allowed to run to completion describes a *simplification hierarchy* with the original model in its leaves, and simplified versions of the model in its interior nodes.

There are many simplification algorithms that fit this paradigm, and they use a variety of primitive simplifications. The algorithm described by Schroeder, Zarge and Lorenson in [17] removes one vertex at a time. Algorithms that remove (collapse) an edge in each primitive simplification include those by Hoppe [9], Ronfard and Rossignac [15] and Garland and Heckbert [4]. Hamann's algorithm [7] removes one face at a time. Brodsky and Watson remove vertex clusters [1].

Simplification algorithms that do not describe and traverse a simplification hierarchy during simplification cannot easily be used with semisimp, though they might conceivably produce hierarchy by being chained to simplify previously simplified output. These include vertex clustering algorithms by Rossignac and Borrel [16] and Low and Tan [13], face merging algorithms [8,11], Turk's retiling algorithm [18], and the simplification envelopes algorithm [3]. Some semisimp functions make use of certain components of the greedy search paradigm (see below for details). In particular, semisimp makes use of the simplification filter and error measure, and manipulates the primitive simplification queue. Adapting semisimp for use with automatic simplification algorithms lacking one or more of these components would require finding or generating substitutes for these paradigm components.

We are aware of only one other tool for semiautomatic simplification, the Zeta tool from Cignoni, Montani, Rocchini and Scopigno [2]. Zeta saves the order in which primitive simplifications are performed, and allows users to manipulate that order. Since Zeta does not make use of simplification hierarchy, there are few order manipulation constraints. However, lacking hierarchy, Zeta does not support propagated geometry manipulation, nor allow segmentation of the simplification process by model region through hierarchy manipulation.

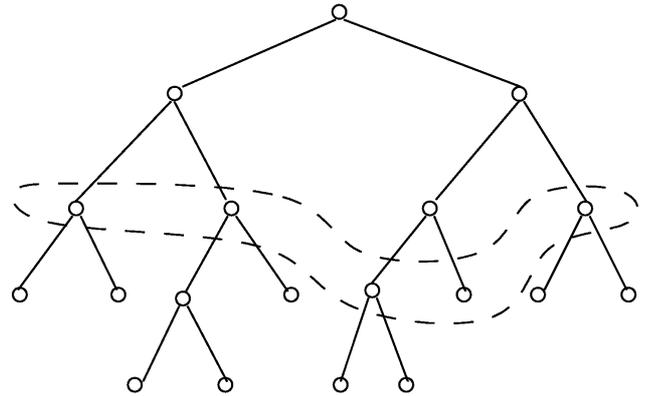

**Figure 1:** A cut across a simple simplification hierarchy.

Because it enables users to manipulate geometry in the multiresolution context of the simplification hierarchy, semisimp is related to research on multiresolution modeling [12,20]. However, while the focus of multiresolution modeling is editing of the original model with control of scale, our focus is the improvement of simplified versions of the original model. Geometric manipulations made with semisimp will typically be much more minor than edits made with multiresolution modelers. Functions unique to our simplified focus include user control of detail distribution through order manipulation, attenuation of geometric manipulation propagation, and partitioning control with hierarchy manipulation.

## 5 semisimp

semisimp is a semiautomatic simplification tool that allows modelers to intervene manually in the simplification process. Users begin by loading the model to be simplified into semisimp, at which point an automatic simplification algorithm is applied. Users can then inspect the results of this algorithm at all levels of detail, improve them where appropriate, and save the results to a file. Saved results may be either a discrete level of detail, or the entire simplification hiearchy.

We implemented semisimp in a Linux environment with OpenGL and Motif libraries. We currently use qslim [4] to perform automatic simplification. The version of qslim we apply considers mesh boundaries and vertex attributes such as texture coordinates and normals during simplification [5]. Input models need not be manifold or closed, though semisimp works most effectively when models are manifold. Our current implementation of semisimp uses the edge collapse as its primitive simplification, and so works best when models are largely topologically connected.

Below we describe in detail the order, geometry and hierarchy manipulations which allow users to implement model simplification improvements. We begin with a brief review of terminology and data structures.

### 5.1 Terminology and Data Structures

semisimp saves and uses as its core data structure the hierarchy created by automatic simplification algorithms using the greedy search paradigm. A model that approximates the original model in its entirety describes a *cut* across the simplification hierarchy (see Figure 1). Each hierarchy node in a cut represents a portion of the original model, called a *patch*. Taken collectively, all the nodes in a cut describe a *partitioning* of the original model.

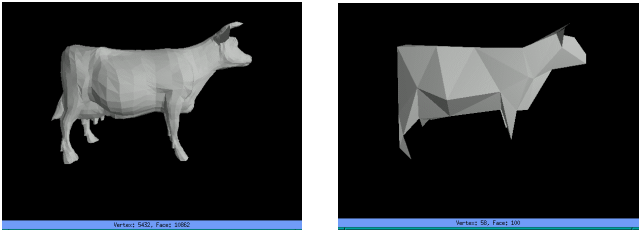

**Figure 2:** Viewing different levels of detail. Here, the cow on the left has 10,000 vertices, the cow on the right 100.

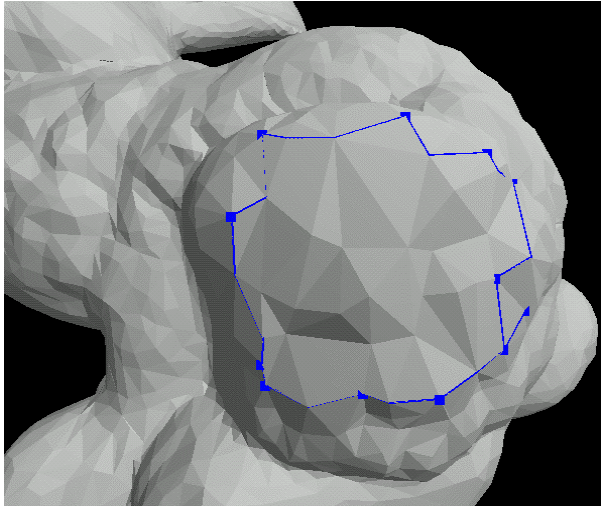

**Figure 3:** Local simplification of the bunny's leg.

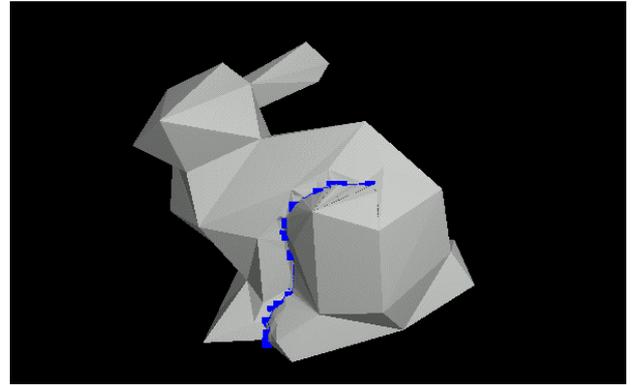

**Figure 4:** Feature preservation along the bunny's leg.

Though the simplification hierarchy is a partial ordering of primitive simplifications (a parent node's simplification must be made after its child's), it is not a complete ordering. semisimp records the order in which simplifications are to be applied in its *order list*. Each element of this list refers to a node in the simplification hierarchy, and the ordering of this list describes a complete ordering of all primitive simplifications. We call the end of the list containing the first several simplifications the *early* end, while we refer to the other end of the list with the last several simplifications the *late* end.

The current level of detail at which the model is viewed is indicated by the *LOD position*. In pointing at the $k$th element of an order list, the LOD position indicates that the first $k$ primitive simplifications should be performed. Each possible LOD position corresponds to a cut across the simplification hierarchy.

## 5.2 Order Manipulation

After the original model is loaded into semisimp, the original model is automatically simplified, and the order list filled. Each element of the order list refers to a node in the automatically generated simplification hierarchy, and the order of those elements mirrors the order in which the automatic simplification algorithm applied its primitive simplifications. Users can navigate through the various levels of detail in the current simplification hierarchy and order list by using a slider to change the LOD position (see Figure 2).

Often a user will be dissatisfied with the distribution of detail provided by the automatic simplification algorithm. For example, on the Stanford bunny, users may wish to exchange a decrease in triangles on the leg for more triangles on the head. semisimp allows users to act on this wish by reordering the elements in the order list. Users can effect a *refining reordering* by moving a primitive simplification to a later position in the order list, delaying the simplification. Conversely, users can effect an *simplifying reordering* by moving a list element to an earlier order list position, performing the simplification more promptly.

In adjusting positions of order list elements, care must be taken to maintain the partial ordering of primitive simplifications defined by the simplification hierarchy. During a refining reordering, a list element $c$ is moved from early position $i$ to late position $k$, where $i < k$. If the parent $p$ of $c$ is located at the position $j$, where $i < j < k$, it is moved to position $k+1$ in the order list. Similar actions are taken for any other ancestors of $c$ found in the range $(j,k)$. During a simplifying reordering, if the child $c$ of the relocated order list element $p$ is found between starting position $k$ and new position $i$, it is relocated to position $i$-1. Similar actions are taken for any other children or descendents of $p$ found in the same range.

semisimp offers the user several interaction techniques for accomplishing both refining and simplifying reordering. With *local simplification* and *local refinement*, users can move the parents or children of a node visible in the current cut to the current LOD position. While viewing the current cut, users highlight a single vertex, a series of edges, or a patch. They then indicate that they would like the highlighted nodes(s) simplified (or refined). All highlighted nodes are visually replaced in the current cut with their parents (children) in the simplification hierarchy. In the case of local simplification, other unhighlighted nodes may also be replaced along with their highlighted siblings. Local simplification is illustrated in Figure 3.

With *feature preservation* and *feature elimination*, users can move the node(s) visible in one LOD position to another LOD position. Users highlight one or more visible nodes in the current cut as above, and then navigate to a different LOD position and indicate that they would like these nodes to be visible there. For feature preservation, users move from an early, detailed LOD position to a later, simplified one. For feature elimination, users move from a late LOD position to an earlier one. Feature preservation is presented (exaggerated for illustration) in Figure 4.

## 5.3 Geometric Manipulation

When viewing a certain cut across the simplification hierarchy, semisimp allows users to improve the position of vertices in the cut using the mouse. To maintain smoothness in the current cut and across different levels of detail, changes in position can be propagated to topological neighbors in the current cut, as well as to ancestors and descendants of the affected nodes in the cut. We should note that `semisimp` is not a full multiresolution modeling tool, and need only support fairly minor edits designed to make a simplified model look more like the original it represents. Therefore the filtering and smoothing techniques we use here are fairly modest, and do not include more advanced filtering and fairing techniques like those described in [20] and [12].

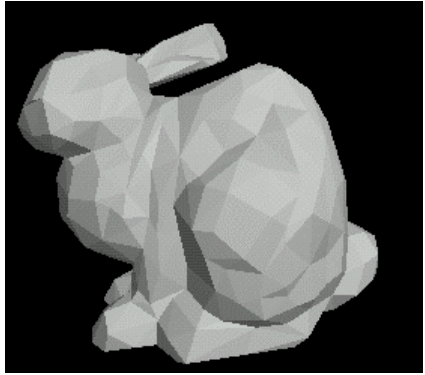

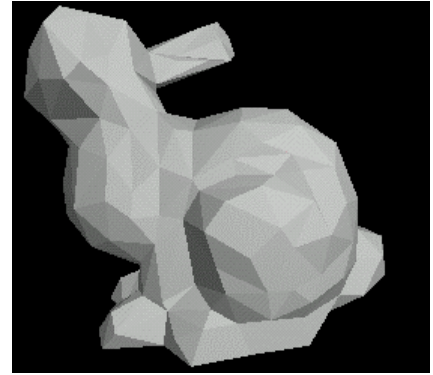

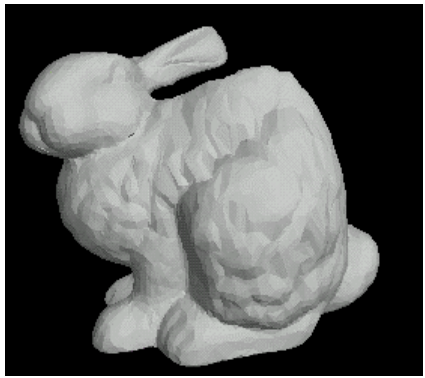

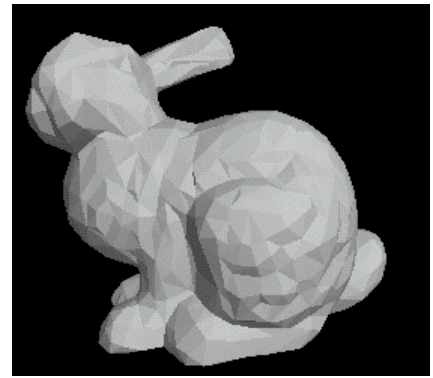

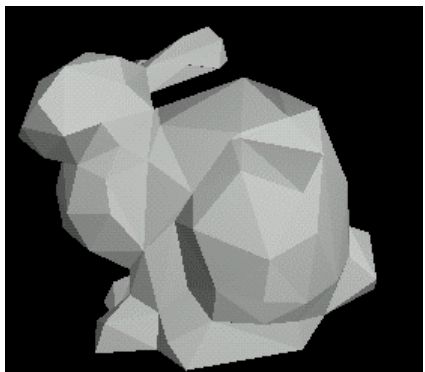

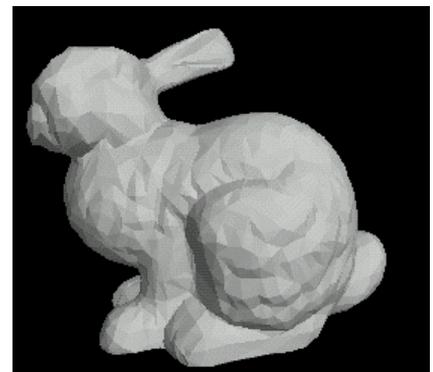

**Figure 5:** The combined effects of neighbor, descendant and ancestor propagation during geometric manipulation. At the top is the manipulated level of detail, followed by affected higher and lower levels of detail.

**Figure 6:** The effects of geometric propagation to children (finer levels of detail) with attenuation. As the detail becomes finer, the manipulation fades away.

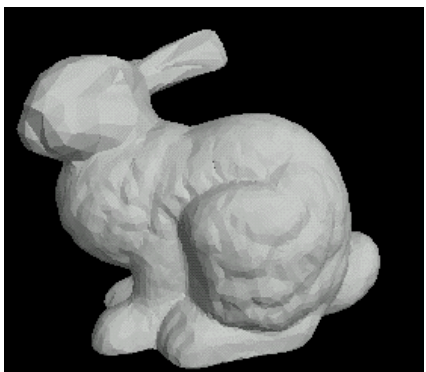

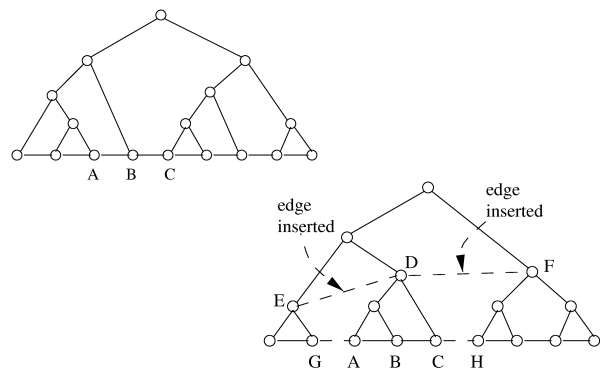

**Figure 7:** The head of the bunny is made a separate partition and subtree, then order manipulation simplifies the head without any blurring at the neck.

**Figure 8:** An illustration of hierarchy manipulation. The user identifies nodes A B and C as a new partition, resimplification then begins to generate a subtree over the partition.

Propagation to topological neighbors of the manipulated vertex $m$ is accomplished by simply interpolating the vector of position change $\vec{\delta}$ across a topological circle with user defined radius $r$ (vertices within $r$ edges of the repositioned vertex are affected by the change). Users can control the shape of the propagated change by manipulating a Bezier curve B defined on the number of edges, with $B(0) = 1$ and $B(r) = 0$. Thus the vector of change $\vec{\delta_i}$ for a vertex $i$ edges away from the manipulated vertex is: $\vec{\delta_i} = B(i)\vec{\delta}$.

To propagate position change to hierarchical ancestors of $m$, we use the simplification filter implemented by the automatic simplification algorithm (see section 4). The simplification filter in `qslim` is the quadric, the sum of the squared plane equations of the faces being summarized, which is minimized to find an approximating vertex position (see [4] for details). We redefine the quadric $Q_m$ of $m$ using the planes of the surrounding simplified faces. The quadric $Q_p$ minimized to find the adjusted position of the parent $p$ of $m$ is then recalculated by once again summing the quadrics of $p$'s children (including $Q_m$). The quadrics of more distant ancestors of $m$ are also recalculated with new sums.

We also provide propagation of the position change to the children of $m$. The propagation can be *direct*, allowing modification of the original model; or *attenuated*, preserving the shape of the original model while allowing some position change at levels of detail close to the manipulated cut.

Direct propagation to descendents is achieved by calculating an orthogonal local coordinate frame for $m$ in its unmanipulated state, and then transforming the global coordinates of $m$'s descendents into detail vectors in the resulting local coordinate space. After $m$ is manipulated, a new local frame is found, and the detail vectors combined with the new frame to generate new global coordinates for $m$'s descendents. We calculate $m$'s frame before and after manipulation by averaging the normal vectors of the faces surrounding $m$ in the current cut to find a vector $\vec{z}$, projecting one of the edges connected to $m$ onto plane orthogonal to $\vec{z}$ to find $\vec{y}$, and setting $\vec{x}$ to $\vec{y} \times \vec{z}$. Thus descendents of $m$ are not only translated, but also reoriented. We considered nested reorientation with calculation of a local frame for every descendent; however, our simplification optimized hierarchies are not balanced and organized into successive frequency bands for multiresolution modeling like those in [20] and [12], making it unclear which faces surround each descendent and frame calculation somewhat arbitrary.

We cast attenuated propagation to descendents as an interpolation problem. If the difference in global position of one of $m$'s descendents $c$ before and after direct propagation is $\vec{d}$, then the attenuated difference in position of $c$ will be $t\vec{d}$, where $0 \le t \le 1$, with $t = 1$ at $m$ and $t = 0$ at any descendent which is a leaf node on the simplification hierarchy. We were tempted to link the value of $t$ to the number of hierarchical generations between $c$ and $m$, but that proved inappropriate, since our simplification hierarchies are usually unbalanced, with leaves having many different depths. We found it most effective to link $t$ to geometry by using the automatic simplification algorithm's simplification error measure (see again section 4). In `qslim` this measure is the value ε returned by the minimization of the quadric used as the simplification filter. Since the quadrics represent the sum of squared distances, if $\varepsilon_c$ and $\varepsilon_m$ are the error measures of the simplifications that produced $c$ and $m$, we set $t = \text{sqrt}(\varepsilon_c/\varepsilon_m)$.

We have found geometric manipulation most effective with all three sorts of propagation: to neighbor, ancestor, and descendant. Figure 5 shows the effects of such a combined propagation. We demonstrate attenuation in Figure 6. Again, in both cases we exaggerate the edit for the purposes of illustration.

## 5.4 Hierarchy Manipulation

To reduce semantic and functional blurring, `semisimp` allows users to halt automatic simplification, correct the current partitioning of the model, and then continue simplification in a segmented fashion. In this way, the head on the Stanford bunny might be simplified separately from the body (see Figure 7). We call this hierarchy manipulation because it changes the structure of the simplification tree by constructing a simplification subtree containing and simplifying the user defined patch. Users gain a new degree of control over both order and geometric manipulation. In the case of order manipulation, hierarchy manipulation sidesteps the constraints of the partial ordering imposed by the automatically defined simplification hierarchy. Hierarchy manipulation improves geometry manipulation by giving users explicit control of propagation to ancestors and descendants.

To manipulate hierarchy, users identify a collection of nodes in the currently viewed cut that should be formed into a new patch and separate subtree $s$ of the hierarchy (see Figure 8). $s$ is formed by recreating the state of the automatic simplification algorithm at the moment it reached the cut (this state may in fact never have been reached if the user has manipulated simplification order), and then modifying that state so that the nodes of $s$ cannot be merged with the remainder of the hierarchy. The automatic simplification algorithm is then restarted and executed until $s$ has been reduced to a single node. At this point, the simplification algorithm's state is again modified to allow the newly formed root of $s$ to be merged with the remainder of the hierarchy, and the simplification algorithm is executed to completion.

`semisimp` recreates simplification state by creating a primitive simplification queue that corresponds to the current cut. With `qslim`, this means the queue contains one primitive simplification entry for each edge in the current cut, sorted by the current simplification error measure. We then ensure that the user defined patch and matching subtree $s$ will not be merged with the remainder of the simplification hierarchy by removing all primitive simplifications crossing the patch boundary from the primitive simplification queue (vertices on the patch boundary are assigned to the patch surround). When `qslim` has simplified the patch to a single node, the primitive simplifications previously removed are reinserted into the primitive simplification queue, minus any that have become redundant during the preceding simplification. Further simplification then incorporates the fully simplified patch.

Care must be taken when using hierarchy manipulation, since it completely rebuilds the higher levels of the simplification hierarchy, making all previous order and geometry manipulations obsolete.

## 6 USAGE EXAMPLES

In this section we present several examples demonstrating the combined application of order, geometric, and hierarchy manipulation with `semisimp`. Figure 9 illustrates the reduction of semantic blurring in a cow model. In the upper row is the original model. The middle row shows the results of automatic simplification with `qslim`. The bottom row shows the results of improvement with `semisimp`. Preservation of the semantically important head and udder has been improved. Users were able to achieve this improvement (and the others discussed in this section) with interactions only at the regions of interest, rather than across the entire model.

Semantic regions may also have an important functional distinction in the target application. For example in animation, regions on articulated models are matched to segments of their skeletons. Automatic simplification can blur the boundaries between these regions. Figure 7 illustrates the use of `semisimp` to preserve a boundary that could have both semantic and functional significance: the neck of the bunny. The head and the body are simplified differently around this boundary.

Finally, Figure 10 illustrates the use of `semisimp` to prevent a different kind of functional blurring: the distortion of a visual discontinuity inside a texture. Such a discontinuity is difficult to detect automatically from model attributes. The first part of the figure shows the original horse model with a single texture applied to it. Although the texture covers the entire model, there is an oval spot in the middle of the texture. After automatic simplification (with texture coordinate preservation), the spot is somewhat distorted. The final part of the figure shows the results after semiautomatic improvement of the simplification with `semisimp`. The oval spot is well preserved.

## 7 FURTHER RESEARCH

Many improvements and extensions of `semisimp` are possible. In particular, geometric manipulations propagated to descendants can easily introduce discontinuities in the model surface when the manipulations deviate significantly from the shape of the input model. It may be possible to reduce these discontinuities with more advanced filtering and smoothing schemes. Propagated geometric manipulations can also alter previously made geometric manipulations. A more elaborate interpolation scheme between manipulated nodes of the simplification hierarchy might be able to solve this problem. More complex editing facilities allowing insertion and deletion of vertices and perhaps editing of non-geometric attributes would be a valuable extension. Finally, although semiautomatic simplification of extremely large models was not our goal, optimization of `semisimp` to handle larger models would be useful. With our current implementation, certain operations (e.g. changing the LOD position) can take several seconds when input models contain several tens of thousands of polygons.

## 8 CONCLUSION

We have presented `semisimp`, a tool for the semiautomatic simplification of highly detailed models. This tool allows users to improve the quality of aggressively simplified models by manipulating the order in which primitive simplifications are applied, the vertex positions of simplified models, and the hierarchical partitioning of the model formed during simplification. With these abilities, users can manually preserve semantically and functionally distinct model regions that are blurred by automatic simplification algorithms, including facial details, regions bound to articulated skeletons, and details embedded in texture mapped images.

## 9 ACKNOWLEDGEMENTS

Dr. John Buchanan, Director of Advanced Technology at Electronic Arts, provided inspiration, financial support and 3D models for this research. The Stanford bunny was provided through the courtesy of the Stanford computer graphics lab.

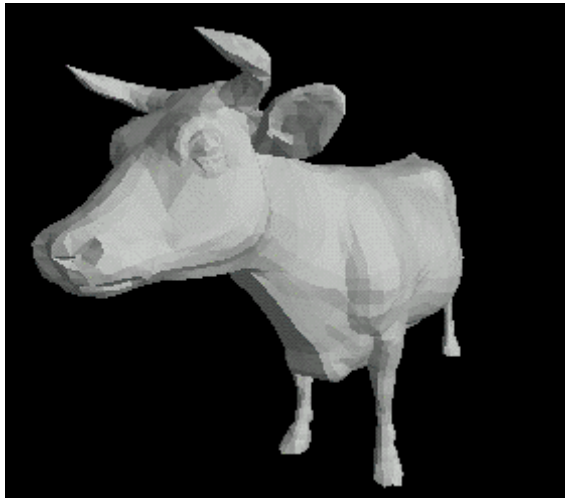
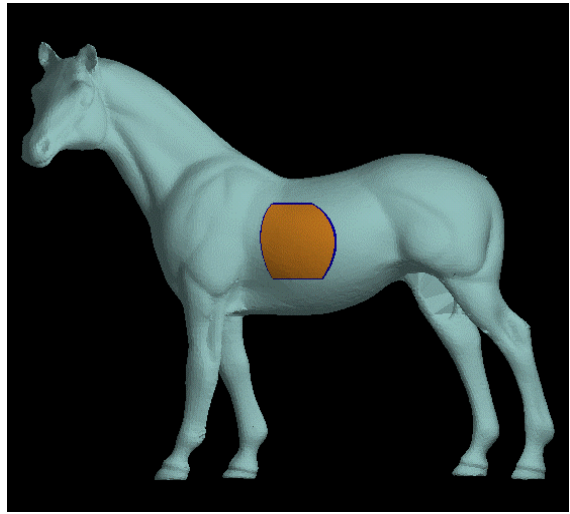

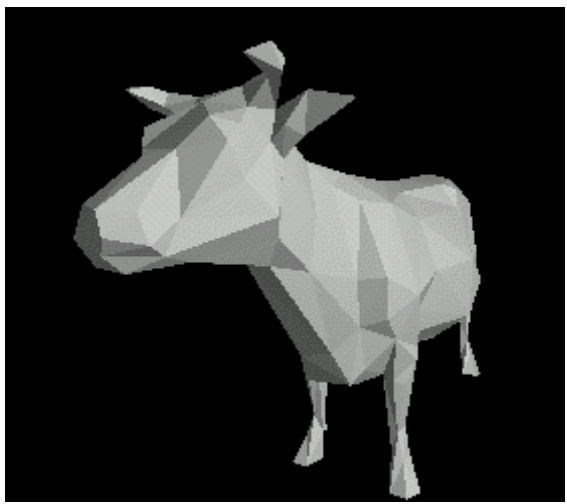
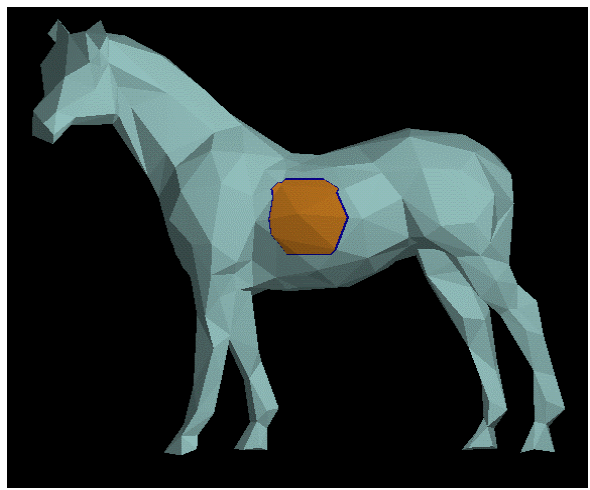

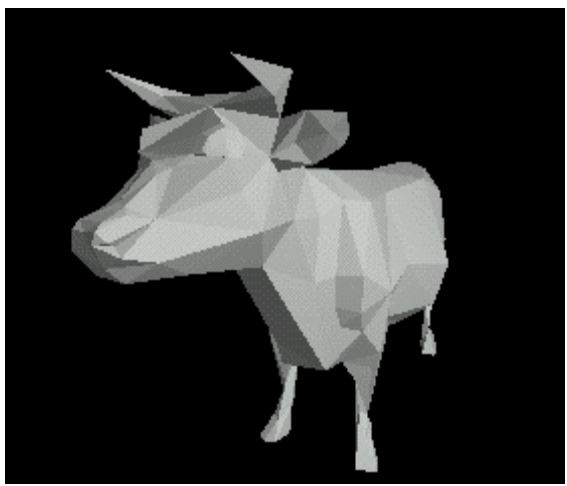
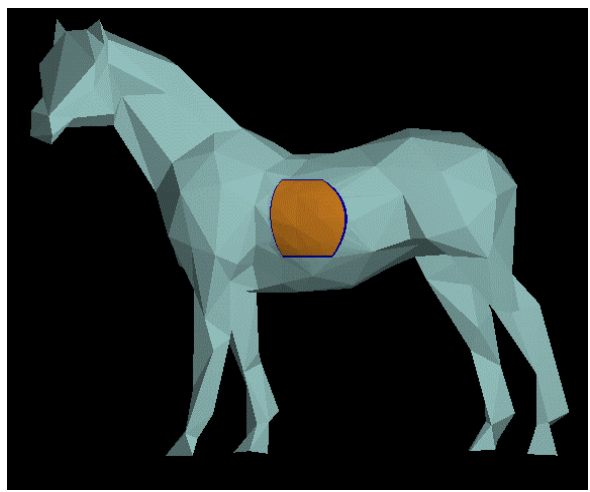

**Figure 9:** Using the combined functions of `semisimp` to reduce semantic blurring of the head. On the top is the original cow. In the middle, the automatically simplified cow with 588 faces. At the bottom, the manually improved cow, also with 588 faces. Notice the strong similarity of the bottom and top models.

**Figure 10:** Using the combined functions of `semisimp` to reduce functional blurring. Here, the entire horse is covered with texture, but there is a strong color discontinuity in the texture. The two lower models have the same number of faces, with the middle produced by `qslim`, the bottom with `semisimp`.